\documentclass[12pt]{article}
\pdfoutput=1
\usepackage{amsmath,amssymb}
\usepackage{epsfig,euscript,array}
\usepackage{amssymb}
\usepackage{graphics}
\usepackage{graphicx}

\newcommand{\tr}{{\rm tr}}

\newcommand{\Ph}{\Phi}

\def\b{\beta}
\def\g{\gamma}

\def\d{\delta}

\def\beq{\begin{equation}}
\def\eeq{\end{equation}}
\def\beqn{\begin{eqnarray}}
\def\eeqn{\end{eqnarray}}
\def\ba{\begin{eqnarray}}
\def\ea{\end{eqnarray}}

\def\m{{\tt -}}

\def\l{\langle}

\def\xprim2bar{\overline{x}^{\prime\prime}}

\def\beq{\begin{equation}}
\def\eeq{\end{equation}}
\def\tr{{\bf tr}}

\newcommand{\beqa}{\begin{eqnarray}}
\newcommand{\eeqa}{\end{eqnarray}}

   \let\b=\beta   \let\g=\gamma   \let\d=\delta
         
      \let\l=\lambda  \let\m=\mu
\let\n=\nu

\let\F=\Phi           
   \let\b=\beta   \let\g=\gamma   \let\d=\delta
         
      \let\l=\lambda  \let\m=\mu
\let\n=\nu

\let\F=\Phi           
\newcommand{\be}{\begin{equation}}
\newcommand{\ee}{\end{equation}}
\newcommand{\bea}{\begin{eqnarray}}
\newcommand{\eea}{\end{eqnarray}}

\def\tr{{\rm tr}\,}
\newcommand{\eq}[1]{Eq.~(\ref{#1})}
\newcommand{\fig}[1]{Fig.~\ref{#1}}

\newcommand{\sect}[1]{Section~\ref{#1}}

\def\A5{(A_5)_{\rm lat}}
\def\thintablerule{\hrule height0.4pt}
\begin{document}

\vbox{\vskip0.0cm}
\begin{flushright}
\vskip 0.7cm
\small{
WUB/13-18
}
\end{flushright}

\vskip 1.0cm
\centerline{\LARGE Non-perturbative Gauge-Higgs Unification:}
{\LARGE\centerline  {Symmetries and Order Parameters.}}

\vskip 2 cm
\centerline{\large Nikos Irges$^1$ and Francesco Knechtli$^2$}
\vskip1ex
\vskip.5cm
\centerline{\it 1. Department of Physics}
\centerline{\it National Technical University of Athens}
\centerline{\it Zografou Campus, GR-15780 Athens, Greece}
\vskip .4cm
\centerline{\it 2. Department of Physics, Bergische Universit{\"a}t Wuppertal}
\centerline{\it Gaussstr. 20, D-42119 Wuppertal, Germany}
\begin{center}
{\it e-mail: irges@mail.ntua.gr, knechtli@physik.uni-wuppertal.de}
\end{center}
\vskip 1.5 true cm
\thintablerule
\vskip 2.0ex
\leftline{\bf Abstract}
\vskip 1.0ex\noindent

We consider pure $SU(N)$ gauge theories defined on an orbifold lattice, 
analogous to the $S^1/\mathbb{Z}_2$
gauge theory orbifolds of the continuum, which according to the perturbative 
analysis do not have a Higgs phase.
Non-perturbatively the conclusion for $N$ even is the opposite, 
namely that spontaneous symmetry breaking does take place
and some of the gauge bosons become massive. 
We interpret this new, non-perturbative phenomenon both mathematically and 
physically.

\vskip 2.0ex
\thintablerule

\vskip-0.2cm
\newpage

\section{Introduction}

In this work we argue that the mechanism of Spontaneous Symmetry Breaking (SSB) in a five-dimensional pure gauge theory is
related to the ability of the system to be sensitive to its global symmetries. Our motivation comes from the finite temperature
deconfinement phase transition. 
With periodic boundary conditions the system is symmetric under a
transformation by a center element.\footnote{
Center transformations in the continuum are non-periodic
gauge transformations $\Lambda(x+L\hat{N})=z\Lambda(x)$, where $z$ is an
element of the center $\mathbb{Z}_N$ of $SU(N)$ and $L$ is the size of the
periodic dimension $N$.} The nature of a certain
order parameter -- the Polyakov Loop, a gauge invariant loop winding one of 
the dimensions -- that transforms non-trivially under this symmetry,
determines the action that can force the system to become aware of its center symmetry: reducing the size of the dimension.
The shrinking of a dimension is an external action to the gauge theory, in the sense that by itself a gauge theory
does not spontaneously change the sizes of its dimensions. Once however this is imposed on it, 
the system at some point responds by undergoing a phase transition.
Without the possibility of breaking the center symmetry and an associated order parameter that controls the breaking,
one would never be able to tell that it is the center symmetry that governs the confinement-deconfinement phase transition.

Apart from the center symmetry (and the global subgroup of gauge transformations) 
the other global symmetries that gauge theories possess originate from the automorphisms
of their local gauge group. Without any external action these symmetries remain inert in the sense that they do not have any
measurable physical consequences.
We will consider a special class of models where the external action involves a projection
of the underlying geometry but also a projection of the algebra with respect to some of its inner automorphisms. 
More specifically, we require the external action to be such that a) translational invariance be broken along one of the dimensions
and b) the original gauge field be broken into a subset of gauge fields and a subset that can be interpreted
as matter. Clearly, these conditions can not be met in four dimensions, without violating observations.
Thus, the minimal version of these models is realized in five dimensions. 
We show in the following that the system responds to the projections by becoming spontaneously aware of its other
global symmetries, notably of its outer automorphisms, which is physically realized by
the system developing a mass gap in its spin 1 sector.
One of our goals is to try to understand if this purely non-perturbative effect has anything to do with the
Higgs mechanism that we observe in the Standard Model.\footnote{
One may ask why not just consider a Higgs-like scalar coupled to a four dimensional gauge system.
From this point of view it is the gauge hierarchy problem associated with these four-dimensional systems that 
provides motivation to study five-dimensional gauge theories.}

In Gauge-Higgs Unification (GHU) models \cite{GHU} the Higgs field originates from the extra-dimensional 
components of a higher than four dimensional gauge field $A_M$, $M=1,\cdots, d$
(the gauge fields are Lie algebra elements $A_M=i\,A_M^A\,T^A$ with $T^A$ the 
Hermitian and traceless generators of the algebra of the gauge group $G$).
The simplest version of GHU models is five-dimensional ($d=5$) gauge theories compactified on the $S^1/\mathbb{Z}_2$ orbifold.\footnote{
Orbifolds entered the high energy physics world through the seminal work of \cite{StringOrb}. 
In the gauge theory context they appeared later \cite{GaugeOrb1,GaugeOrb2}.}
As a result of the orbifold boundary conditions, the fifth dimension
becomes an interval thus breaking translational invariance, the original five-dimensional 
gauge group $G$ breaks on the four dimensional boundaries at the ends of the interval down to $H$
and some of the extra dimensional components of the gauge field transform as matter under $H$ -- the candidate for a Higgs 
with perturbatively finite mass \cite{Kubo,Quiros}.
This is the external action on the system the spontaneous respond to which we intend to study,
in the spirit of finite temperature phase transitions. 
It is important to recall that the embedding of the orbifold action in the algebra is 
typically via the rank preserving inner automorphism
\be
A_M \longrightarrow g\,A_M\,g^{-1}\, ,\label{inneraut}
\ee
with $g$ an appropriate element of $G$. 
Inner automorphisms induce transformations that can be always represented as group conjugations.
Actions of the type \eq{inneraut} 
trigger the breaking patterns $G\to H$, with $H$ an equal rank subgroup of $G$. 
For example for $G=SU(N)$
one has $SU(p+q)\longrightarrow SU(p)\times SU(q)\times U(1)$ (see for example \cite{GaugeOrb2}).
The question of our interest then is, under what circumstances $H$ can somehow further break,
resulting in the breaking sequence
\be
G\longrightarrow H\longrightarrow E\label{break}
\ee
with the first, rank preserving breaking due to the orbifold boundary conditions and the second, rank reducing breaking due to SSB.

The perturbative analysis of these models states that if some component of $A_5$ acquires a vacuum expectation value (vev) $v$,
then the 1-loop Coleman-Weinberg potential possesses a non-trivial minimum which breaks $H$ spontaneously to a subgroup $E$,
only if fermions of appropriate representations and boundary conditions are coupled to the gauge field. 
SSB realized in this way is called the Hosotani mechanism. According to perturbation theory, in the pure gauge theory the second 
stage in \eq{break}, that of the spontaneous breaking is therefore absent.
Let us see what happens non-perturbatively.

\section{Global symmetries, phases  and order parameters}

\subsection{The periodic lattice \label{sec:tor}}

The first thing one would like to understand is the general structure of the phase diagram.
Let us consider for a moment a five-dimensional, infinite, periodic lattice with a pure gauge theory with local symmetry $G$ defined on it.
Gauge links in direction $N$ at the node $m_M$ are denoted as $U_N(m_M)$. There are $L^5$ nodes in the lattice.
The phase diagram can be split, to begin, at most into two types, the confined and the deconfined phase.
The process in order to distinguish these two phases is already described in the Introduction and here we reiterate it, adjusted this time to the lattice.
One typically proceeds by identifying a global symmetry of the lattice action that is not a gauge transformation and an order parameter
that transforms non-trivially under it. In a theory without fundamental scalar fields and with periodic boundary conditions the global symmetry 
is $Z: U_N\to zU_N$ at a fixed slice orthogonal to direction $N$, such that $z$ lies in the center of $G$. 
That $z$ is a center element guarantees that under the transformation a group element remains a group element ($Z$ should not break $G$), the action is invariant
and it is not a gauge transformation since under $Z$, links do not transform covariantly.
A gauge invariant order parameter that transforms non-trivially
is the Polyakov Loop $P$
\be
P = \prod_{m_M=0}^{L-1} U_N(m_M)\, ,
\ee
a loop that winds the dimension $N$ of the lattice: $Z: P\to zP$. The external action necessary to expose $Z$ is
reducing the number of lattice nodes in the $N$-direction.
Then, the confined phase is defined as the phase where $\langle P \rangle = 0$ and the deconfined phase
where $\langle P \rangle \ne 0$. Monte Carlo analysis of the phase diagram of the five-dimensional periodic $SU(2)$ 
theory can be found in \cite{TorusMC2}.

Let us now imagine that we are in the deconfined phase and ask if we can further characterize it as a Coulomb
or as a Higgs phase.\footnote{ By Higgs we mean here a strictly spontaneously broken phase 
where some of the gauge bosons become massive. This excludes from our discussion
the mass that gauge bosons may acquire from magnetic monopoles 
\cite{'tHooft}, \cite{Shigemitsu}.} 
Following the previous line of thought, a Higgs phase exists if and only if
a gauge invariant order parameter that transforms non-trivially under a global, non-gauge symmetry, the breaking of which can trigger the breaking of $G$,
takes a non-zero expectation value. The first task then is to find such a global symmetry and then the corresponding order parameter.
The automorphism group of $G$ pertains on the lattice so we have a candidate for the global symmetry.
Regarding the order parameter, since $\tr(P)$ is invariant under automorphism group transformations, a new order parameter is needed.
The operator that can play this role has the generic form
\be
O = -iT^A \, V \label{Z}
\ee
with $V$ a gauge invariant object that can be arranged to have the quantum numbers of a vector boson.
However, ${\tr } \{O\}$ is not gauge invariant for non-Abelian groups.
The reason is that products of adjoint representations, never contain a fundamental representation
and in order to make $V$ in \eq{Z} gauge invariant one needs at least one object in the fundamental representation \cite{Hart}.
We conclude that in this case since there is no external action that can expose the inert global symmetries
and (consistently) no associated order parameter, SSB can not be realized in the periodic, pure gauge system. 
The deconfined phase must be purely Coulomb. 
Next we turn to the orbifold lattice.

\subsection{The orbifold on the lattice \label{sec:orb}}

We first repeat the properties of the orbifold lattices necessary to study GHU non-perturbatively, following their construction in \cite{LatOrb1}.
Consider lattices of dimensionless size $L^4$ in the four-dimensional sense and $N_5$ in the fifth-dimension. 
We will be often taking $L\to \infty$ but we will always keep $N_5$ finite.
The nodes of such a lattice are denoted by $n_M=\{n_{\m}, n_5\}$ with $\m=1,\cdots,4$ and $n_5=0,\cdots , N_5$.
The orbifold boundary conditions are implemented in the gauge group via an $SU(N)$ element $g$, such that $g^2$ is in the center of $SU(N)$.
The action of $g$ on the lattice links is via the inner automorphism 
\be
U(n_M,N) \longrightarrow g\,U(n_M,N)\,g^{-1}\, .
\ee
Only gauge transformations that commute with $g$ are allowed on the boundaries. 
In other words, $g$ is an element of $C_G(H)$, the centralizer\footnote{
The centralizer of a subgroup $H$ of $G$ is defined as
$C_G(H) = \{ g_G \in G \, |\, h g_G = g_G h \;\forall h\in G \}$.}
of $H$ in $G$.
Because of this, the lattice links
have to be split in three types: links on the "left" (right) boundary $U(n_\m, n_5=0; \n)\equiv {\cal T}_\n(n)$ ($U(n_\m, n_5=N_5; \n)\equiv {\cal V}_\n(n)$), 
links along the extra dimension $U(n_\m, n_5;5)\equiv {\cal U}(n_5)$ and the rest, to which we do not assign any special notation.
We will generally refer to ${\cal U}(0)$ and ${\cal U}(N_5-1)$ as "hybrid" links. 
The proper gauge transformations for the lattice orbifold are
\bea
&& {\cal T}_\n(n) \longrightarrow 
\Omega_H(n)\, {\cal T}_\n(n)\, \Omega_H(n+\hat{\n})^\dagger \,, \quad  
{\cal V}_\n(n) \longrightarrow 
\Omega_H(n)\, {\cal V}_\n(n)\, \Omega_H(n+\hat{\n})^\dagger \,,
\nonumber\\
&& {\cal U}(0) \longrightarrow 
\Omega_H(0)\, {\cal U}(0)\, \Omega_G(1)^\dagger \,, \quad 
{\cal U}(N_5-1) \longrightarrow 
\Omega_G(N_5-1)\, {\cal U}(N_5-1) \Omega_H(N_5)^\dagger \,,
\eea
and for all other links $U_M(n)\longrightarrow \Omega_G(n)\, U_M(n)\, 
\Omega_G(n+\hat{M})^\dagger$.
Here $\Omega_H\in H$ with $[g, \Omega_H]=0$ and $\Omega_G\in G$.
The set of gauge transformations given above define the local symmetry ${\cal G}$ of the lattice orbifold.
In the following, when an operator or a transformation property depends on a single space-time dummy index, the index will be sometimes suppressed.
The ${\cal G}$ invariant action we use will be generally anisotropic, with $\b_4$ a coupling
multiplying all four-dimensional plaquettes $U_4(p)$ and $\b_5$ multiplying plaquettes with two sides along the fifth dimension $U_5(p)$:
\bea
S[U] &=& \frac{1}{N} \sum_{\rm 4d-plaq.} \b_4\, 
\left(1-\frac{1}{2}\d_{n_5,0}\right)\,\left(1-\frac{1}{2} \d_{n_5,N_5}\right)\,
{\rm Re}\,\tr\,\left[1-U_4(p)\right] \nonumber\\
&+& \frac{1}{N} \sum_{\rm 5d-plaq.} \b_5\,
{\rm Re}\,\tr\,\left[1-U_5(p)\right] \,. \label{action}
\eea
Note that only plaquettes with a counterclockwise orientation are summed over.
The isotropic lattice is realized for $\beta_4=\beta_5$. 
The above defines what we call from now on the orbifold lattice.
Notice that no boundary terms are required in \eq{action}.
For a more detailed description see \cite{LatOrb1}.
Notice that on the orbifold lattice, the breaking pattern we are interested in is expressed as
\be
{\cal G} \longrightarrow {\cal H} \longrightarrow {\cal E}\, .
\ee
We use calligraphic letters for the lattice local gauge symmetries because they are realized in a particular way, mainly due to the hybrid links.
As a group though, ${\cal H}$ is isomorphic to $H$.

We define the left-to-right boundary-to-boundary-line 
\be
l = \prod_{n_5=0}^{n_5=N_5-1}{\cal U}(n_5)\label{line}
\ee
transforming as $l\rightarrow \Omega_H(n_5=0)\, l\, \Omega_H(n_5=N_5)^\dagger$ under ${\cal G}$,
and from it the orbifold projected scalar Polyakov Loops $P_L$ and $P_R$
\bea
P_L &=& l\, g\, l^\dagger\, g^\dagger \,, \label{PL} \\ 
P_R&=& l^\dagger\, g^\dagger\, l \, g \,. \label{PR}
\eea
$P_L$ can be thought of as a field living on the left boundary and $P_R$ as
a field on the right boundary. 

Scalar operators can be defined as $\tr(P_{L(R)})$ or as $\tr(\Phi^\dagger\Phi)$
using for $\Phi$ one of the expressions
\be
\Phi_{L(R)} = \frac{1}{4N_5} [P_{L(R)} - P_{L(R)}^\dagger, g].\label{Phi}
\ee
These operators were introduced in \cite{OrbifoldMC}.

We distinguish two types of vector boson operators, for which
we use the symbol $Z_k$ with spatial index $k=1,2,3$.
The first type has the same building blocks
as the Polyakov loops in \eq{PL} or \eq{PR} but there is only one
insertion of $g$.
The $Z$-operator introduced in \cite{OrbifoldMC} (inspired by
\cite{Montvay}) and defined 
on the left boundary is
\be\label{eq:ZL}
\left.Z_{Lk}(n)\right|_{n_5=0} = 
g\,{\cal T}_k(n)\,\F_L(n+\hat{k})\,{\cal T}_k(n)^\dagger\,\F_L(n) 
\,.
\ee
Analogously we can define a $Z$-operator on the right boundary
\be\label{eq:ZR}
\left.Z_{Rk}(n)\right|_{n_5=N_5} = 
g\,{\cal V}_k(n)\,\F_R(n+\hat{k})\,{\cal V}_k(n)^\dagger\,\F_R(n) \,.
\ee
$\tr(Z_{L(R)k})$
are vector operators of spin 1, have parity $P=-1$ and charge conjugation 
$C=-1$, see Appendix~\ref{sec:app_transf}.
The gauge invariance of $\tr(Z_{L(R)k})$
relies on the fact that $g$ commutes with any $H$ gauge transformation
and since the centralizer
$C_{SU(p+q)}(SU(p)\times SU(q)\times U(1))\equiv \mathbb{Z}_{p+q}\times U(1)$ \cite{Patera},
it is unlikely that other, independent operators of this type can be constructed.

A second type of $Z$-boson operators can be constructed
using the operators listed in \cite{Yaffe}. We define
\begin{eqnarray}\label{eq:Zpm}
Z^+_{L(R)k}(n) & = & 
\Phi_{L(R)}(n)\,\{\hat{F}_{12}(n),\hat{F}_{k5}(n)\} \,, \\
Z^-_{L(R)k}(n) & = & 
\Phi_{L(R)}(n)\,[\hat{F}_{12}(n),\hat{F}_{k5}(n)] \,,
\end{eqnarray}
where $n_5=0$ ($n_5=N_5$) for the operators on the left (right) boundary.
The lattice expression for the field strength tensor $\hat{F}_{MN}$ is given 
in \eq{eq:F}.
The operators $\tr(Z^\pm_{L(R)k})$ have parity $P=-1$, charge conjugation $C=\pm1$ and
spin $J=1$, see Appendix~\ref{sec:app_transf}.

In Appendix~\ref{sec:app_cl} we show that
both type of $Z$ operators
have the same trace structure thus they contain the same
spectrum of gauge bosons.

\subsection{Global symmetries of the orbifold \label{sec:globalsym}}

We discuss here only the symmetries that are not in the global subgroup of gauge transformations.
Given this premise, by examining the action \eq{action} we find the global symmetry
\be\label{eq:globalsym}
Z\times F\times {\rm \bf Aut} \, .
\ee
$Z$ is the transformation by a center element of ${G}$ and governs the phase transitions on four-dimensional hyperplanes.
$F$ is the reflection symmetry around the middle of the fifth dimension. It is a non-local symmetry as it relates for example
the two boundaries.

${\rm \bf Aut}$ is the group of automorphisms of $H$. It consists of the elements that
descend from the automorphism group of $G$ including "accidental" elements
such as outer automorphisms related to the interchange of two identical group factors in $H$.
An example of an accidental automorphism of $H$ is met in the $SU(4)\stackrel{g}\longrightarrow SU(2)\times SU(2)\times U(1)$ model.
Automorphisms induced by $G$ on $H$ also contain the non-accidental outer automorphisms of $H$. 
The latter can be identified with the charge conjugation
operator $C$ for any $SU(N)$ group with $N\ne 2$, including the $U(1)$ case. 
On the lattice charge conjugation acts as complex conjugation of the
gauge links, see Appendix~\ref{sec:app_C}.
In other words, at the level of the Lie algebra charge conjugation acts as
\begin{equation}\label{eq:out}
T^a \;\longrightarrow\; -(T^a)^* \,, 
\end{equation}
which is an outer automorphism of the Lie algebra.
The only special case is $SU(2)$ which has no outer automorphisms,
since charge conjugation is equivalent to a global gauge transformation by
$(-i\sigma^2)$.
Two key properties that we note are that in general projecting a gauge theory by outer automorphisms
induces the breaking of its rank and that charge conjugation, when associated
with an outer automorphism (i.e. for all $SU(N)$ except $SU(2)$) in general can not be represented as a group conjugation.
Finally, an outer automorphism of $H$ in some cases can be represented as a group conjugation when it is an induced 
outer automorphism of $G$ on $H$. 
This will be analyzed in detail in the following.

We start by defining the group of fixed point symmetries
\begin{equation}\label{eq:calF}
{\cal F}={\cal F}_L\oplus {\cal F}_R
\end{equation}
The transformations in ${\cal F}_L$ are defined as
\begin{equation}\label{eq:calFL}
{\cal U}(0) \longrightarrow g_F^{-1}\,{\cal U}(0)\,, \quad
{\cal T}_\n(n_\m) \longrightarrow g_F^{-1}\, {\cal T}_\n(n_\m)\, g_F \,,
\end{equation}
where $g_F$ is a constant matrix in the normalizer of $H$ in $G$,
the group $N_G(H) = \{ g_G \in G \, |\, g_G^{-1} H g_G = H \}$.
Links not included in the subset specified by \eq{eq:calFL} are unchanged.
Analogously the transformations in ${\cal F}_R$ are defined as
\begin{equation}\label{eq:calFR}
{\cal U}(N_5-1) \longrightarrow {\cal U}(N_5-1)\, g_F \,, \quad
{\cal V}_\n(n_\m) \longrightarrow g_F^{-1}\, {\cal V}_\n(n_\m)\, g_F \,.
\end{equation}
The hybrid links ${\cal U}(0)$ (${\cal U}(N_5-1)$)
transform under ${\cal F}_L$ (${\cal F}_R$) like a matter field.
The transformations \eq{eq:calFL} and \eq{eq:calFR} leave separately
the action invariant.

The symmetry transformations in ${\cal F}$ have been introduced in 
\cite{Stick}, where the following argument is presented.
The transformations in ${\cal F}$ have to be consistent with the
orbifold projection. Consider $g_F\in{\cal F}_L$ and $h_i$ is a link
on the left boundary. The following diagram has to be consistent
\bea
h_i & \xrightarrow[]{g_F} & g_F^{-1}\,h_i\,g_F \equiv h_j \in H \nonumber \\
g\,\downarrow & & \downarrow\,g \nonumber \\
g\,h_i\,g^{-1} & \xrightarrow[]{g_F} & X \nonumber
\eea
therefore the quantity $X$ has to satisfy the property
\be
X = g\,\left(g_F^{-1}\,h_i\,g_F\right)\,g^{-1} 
  = g_F^{-1}\,\left(g\,h_i\,g^{-1}\right)\,g_F \,. \nonumber
\ee
It follows that
\begin{equation}\label{eq:zG}
g\,g_F = z_G\,g_F\,g \,,
\ee
where $z_G$ is an element of the center of $G$, i.e. it commutes with
any element of $G$ (and $H$).
The transformations of lattice operators under ${\cal F}_{L(R)}$ are
summarized in Appendix~\ref{sec:app_calF}.

When $z_G$ in \eq{eq:zG}
is equal to the identity $I$, the transformations 
in ${\cal F}$ are either global gauge transformations or transformations
which do not break the rank and are therefore inner automorphisms of $H$.
We are interested in the case $z_G\neq I$, which is an outer automorphism.
In this case, following  \cite{Stick}, we call the transformations in ${\cal F}$ ``stick'' symmetries
and denote $g_F\equiv g_s$, with
\begin{equation}\label{eq:stick}
\{g,g_s\}=0 \,.
\end{equation}
Clearly, in an element $g_s$ we are looking at
an element of the group $W_G(H)=N_G(H)/H$, called the "generalized Weyl group" in \cite{Shankar}.
In particular, one finds that
\bea
W_{SU(n+1)} (SU(n)\times U(1)) & = & \left\{\begin{array}{ll}
                       \mathbb{Z}_2 & \mbox{if $n=1$}\\
                     {\rm trivial} & \mbox{if $n>1$} 
                   \end{array} \right.\label{WS}
\eea
telling us that in the $SU(2)\stackrel{g}\to U(1)$  orbifold model ($SU(1)$ factors in \eq{WS} are redundant) we should expect finding
a stick symmetry, while in the $SU(3)\stackrel{g}\to SU(2)\times U(1)$ orbifold model, we should not.
In fact, in the classified cases, whenever $W_G(H)$ is non-trivial, it is a $\mathbb{Z}_2$ symmetry.
A practical way to recognize cases where a stick symmetry might exist is to look at the 
orbifold projection matrix $g$: a stick symmetry is likely to exist when $\tr g = 0$ \cite{Stick}.
The simplest class of such models is the one with $G=SU(2n)$, with the lattice defined by
generators in the fundamental representation and $g={\rm diag}(1_n, -1_n)$, where
$1_n$ is the $n$-dimensional unit vector. Notice that this class includes non-trivial cases that are not 
contained in \eq{WS}, as it includes also models with an accidental outer automorphism.
Such an example is the $G=SU(4)$ orbifold model on which we elaborate below.
We note another interesting case. 
It is the $Sp(4)\stackrel{g}\to SU(2)\times SU(2)$ orbifold model\footnote{
This, as well as the $G=SU(4)$ model, have been considered as a possible GHU models in \cite{frere}.},
where the non-perturbative SSB mechanism should be at work.
Even though this may not be the most convenient model for Monte Carlo simulations,
it could be interesting from a theoretical point of view.

We are therefore left to consider groups $G=SU(2n)$ with $n\in\mathbb{N}^+$,
which do have a stick symmetry with $z_G=-I$.
The stick symmetry is a global transformation which is not a global
gauge transformation. It can be spontaneously broken,
consistently with Elitzur's theorem \cite{Elitzur}.
Let us denote by $S_L$ and $S_R$ the eigenvalues ($\pm1$) of operators 
under the stick transformations contained in 
${\cal F}={\cal F}_L\oplus {\cal F}_R$. The values of $S_L$ and $S_R$ can 
be found using the results of Appendix \ref{sec:app_calF} 
by inserting $z_G=-I$ (which defines stick transformations).
We are interested in the value of $S=S_L\cdot S_R$ since the product of
the transformations on the left and on the right boundary respect the
reflection symmetry $F$. The operators $\tr(P_{L(R)})$ are even ($S=1$)
whereas the operators $\tr(Z_{L(R)k})$ are odd ($S=-1$).
Therefore a non-zero expectation value of $\tr(Z_{L(R)k})$ breaks spontaneously
the stick symmetry.\footnote{
The operator $\tr(Z_{L(R)k})$ is not a Euclidean invariant. In a simulation one
can measure for example $\sum_k[\tr(Z_{L(R)k})]^2/3$, see \fig{fig:orderpar}.
What is meant here is an effective potential for $\tr(Z_{L(R)k})$, in analogy
with the Standard Model Higgs. 
We will return to this point in \sect{sec:NPGHU}.}
The breaking of the stick symmetry induces the breaking of the group
${\cal F}$, which contains global gauge transformations as well,
meaning that there will be massive gauge bosons.
The deconfined phase becomes a Higgs phase.

The only possibility to break the rank, which is alternative to the
stick symmetry and would be available also for groups $G=SU(2n+1)$, is
through the outer automorphism of charge conjugation $C$.
But this implies that the photon, which has $C=-1$, would become
massive and this ``solution'' for breaking the rank has to be
dismissed.

Now we present explicit examples for the groups $SU(2)$, $SU(3)$ and $SU(4)$.

\subsection{The $SU(2)$ orbifold \label{sec:SU2}}

The case of the $G=SU(2)$ orbifold is the simplest of a class of 
models that have rank reducing automorphisms
that can be expressed as group conjugations and are
amenable to Monte Carlo
simulations \cite{OrbifoldMC,Knechtli:2014ioa,OrbifoldMC2}.
%
\begin{figure}[t]
\begin{center}
   \includegraphics*[angle=0,width=.48\textwidth]{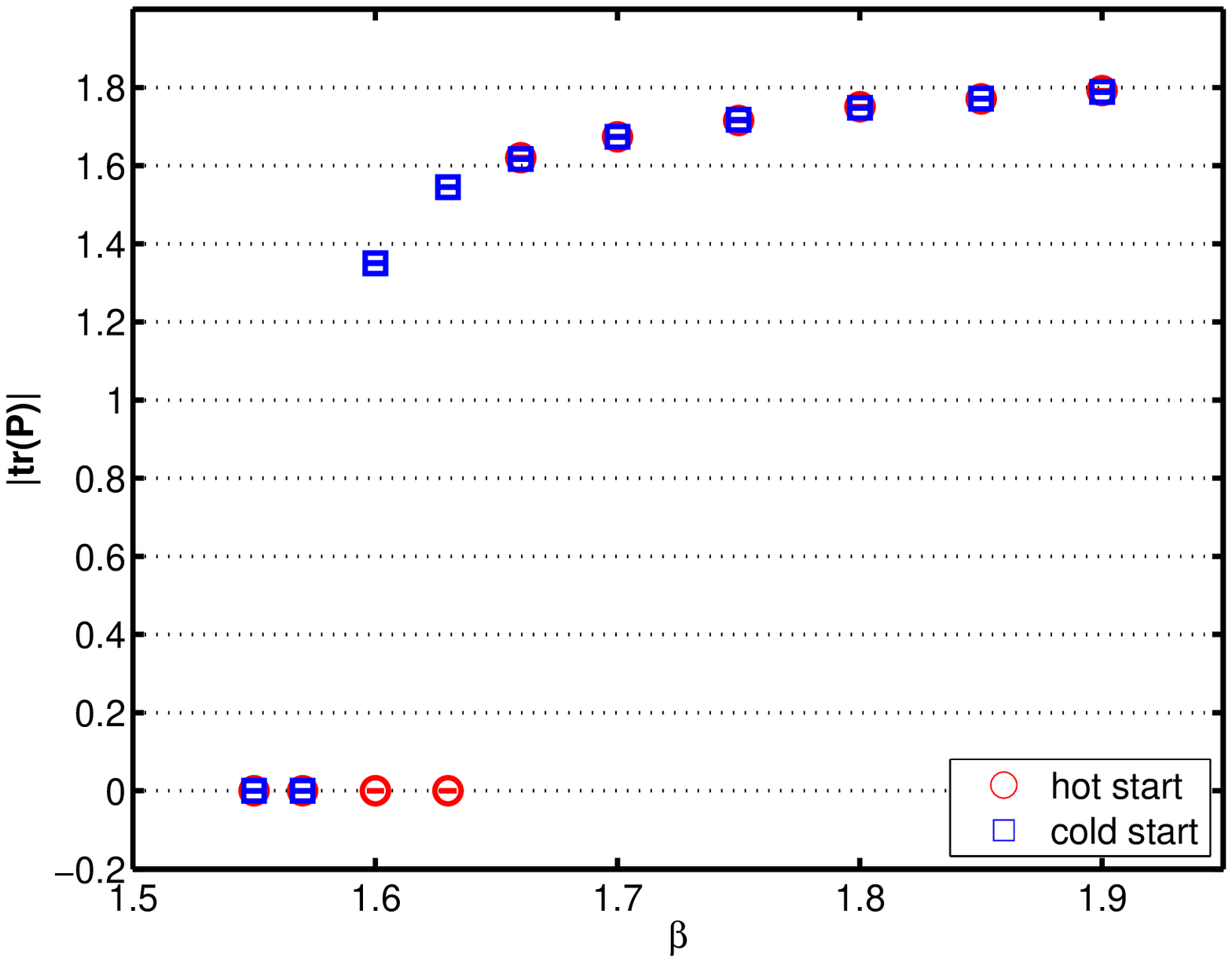}
   \includegraphics*[angle=0,width=.48\textwidth]{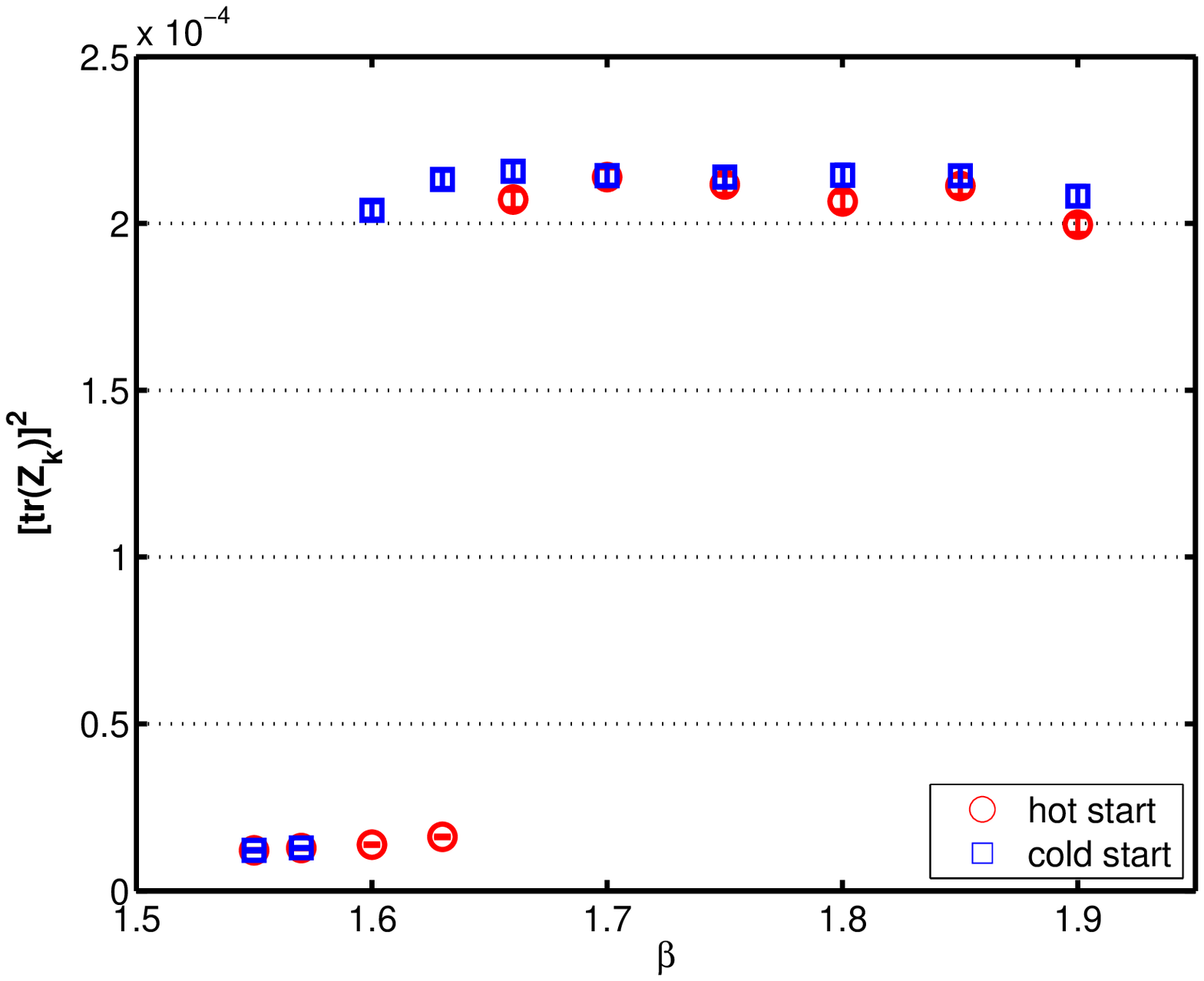}
\end{center}
  \caption{\small Scalar Polyakov loop (left plot) 
and vector Polyakov loop (right plot) from Monte Carlo simulations 
of the $G=SU(2)$ orbifold. Monte Carlo averages from hot and cold 
starts on $24^4\times5$ lattices are shown as a function of 
$\beta=\beta_4=\beta_5$.}\label{fig:orderpar}
\end{figure}
%
In the case of $G=SU(2)$ where $g=-i\sigma^3$, we have 
$H=\{\exp(\omega g) \;,\; \omega\in\mathbb{R}\}=U(1)$.
There is a stick symmetry realized by $g_s=-i\sigma^2$
or equivalently by $g_s=-i\sigma^1$.
In short, in this model we have the breaking pattern
\be
SU(2)\stackrel{g}\longrightarrow U(1)\stackrel{\rm SSB}\longrightarrow 
D \label{SU2break}
\ee
where ${\cal E}=D$ can be either trivial or a remnant $\mathbb{Z}_2$ subgroup of $U(1)$.
We conclude that the rank of $H$ is broken due to the spontaneous breaking of the 
generalized Weyl (or stick) symmetry group,
the only available non-trivial automorphism that the system can access.

\fig{fig:orderpar} shows Monte Carlo results for the quantities
$|\tr(P_L)|$ defined in \eq{PL} (left plot) and
$\frac{1}{3}\sum_k[\tr(Z_{Lk})]^2$ defined in \eq{eq:ZL} (right plot).
The loops $P_L$ and $Z_{Lk}$ are averaged over the points of the 
four-dimensional boundary.
The lattices have size $24^4\times5$ points and in order to locate phase
transitions, results of simulations starting from a hot (random) and 
a cold (identity matrix) gauge link configuration are shown.
A first order transition manifests itself as an hysteresis where the
results from hot and cold start differ.
The statistics of each simulation is 4000 measurements 
separated by two update iterations, each iteration consisting 
of one heatbath sweep and 12 overrelaxation sweeps.
The thermalization is 1000 update iterations.
The scalar and vector Polyakov loops are measured using links smeared by
10 iterations of HYP smearing \cite{Hasenfratz:2001hp} adapted to the orbifold
\cite{YoneyamaPhD}.
Both observables show an hysteresis at values $\beta_c=1.60$--$1.63$
($\beta=\beta_4=\beta_5$) thus confirming the presence of
a first order bulk phase transition (the plaquette has a similar behavior).
The transition is from the confined phase at $\beta<\beta_c$ (where both
observables are zero or close to zero) into the Higgs phase at $\beta>\beta_c$
(where both observables become non-zero). In the latter phase,
the mass of the $Z$ boson can be extracted from correlators of 
$\tr(Z_{Lk})$ and it is found to be non-zero \cite{OrbifoldMC,Knechtli:2014ioa}.
Therefore we call the phase at $\beta>\beta_c$ a Higgs phase.
Because it yields the value of the $Z$ boson mass, we identify the operator 
constructed from $\tr(Z_{Lk})$ as order parameter of the Higgs phase.
In addition, since $\tr(Z_{Lk})$ is odd under the stick symmetry,
the Monte Carlo results verify the breaking pattern in \eq{SU2break}.

Finally we notice that the Monte Carlo results show that the gauge boson
mass is non-zero everywhere for $\beta>\beta_c$ \cite{YoneyamaPhD}.
It diminishes towards the perturbative limit $\beta\to\infty$ where it
is expected to be zero. This means in particular that spontaneous symmetry
breaking is not a lattice (strong coupling) artifact.

\subsection{The $SU(3)$ orbifold \label{sec:SU3}}

Consider the example of $SU(3)$ with the orbifold projection
$g={\rm diag}(-1,-1,1)$ that leaves the symmetry ${H}=SU(2)\times U(1)$ on the boundaries.\footnote{In the continuum 
the orbifold properties of a bulk group that is, or contains $SU(3)$ have been studied in \cite{SU3cont}.}
One can easily check that, as expected from the general group theoretical discussion, there is no $SU(3)$ stick matrix $g_s$.
All there is in the group of fixed point symmetries are transformations which
commute with $g$ and cannot break the rank. 
The non-perturbative gauge symmetry breaking mechanism is absent.

\subsection{The $SU(4)$ orbifold \label{sec:SU4}}

The next example is the $G=SU(4)$ orbifold where if we take $g={\rm diag}(1,1,-1,-1)$
we have $H=SU(2)\times SU(2)\times U(1)$ surviving on the boundaries.
This seems to be the simplest example where $G$ is unitary, the electroweak group can be embedded in $H$
and the Higgs mechanism is realized in a non-perturbative way.

The matrices
\be
g_s  = -i\begin{pmatrix} 
0& 0 & 0 & 1 \cr 
0& 0& 1 & 0 \cr
0& 1& 0 & 0 \cr
1& 0& 0 & 0 \cr
\end{pmatrix}\, \hskip 1cm
g_s'  = -i\begin{pmatrix} 
0& 0 & 1 & 0 \cr 
0& 0& 0 & 1 \cr
1& 0& 0 & 0 \cr
0& 1& 0 & 0 \cr
\end{pmatrix}
\label{stick4}
\ee
fulfill all the constraints that stick matrices are supposed to. The stick symmetry is related
to the accidental outer automorphism due to the interchange symmetry and the internal charge conjugation of the two $SU(2)$ factors.
It is a $\mathbb{Z}_2 \times \mathbb{Z}_2$ transformation 
(it is not a $\mathbb{Z}_4$ transformation since $g_s$ commutes with $g_s'$ and $g_s g_s'$ commutes with $g$) and $z_G^2=I$ in \eq{eq:zG}.
For the transformation of the $SU(4)$ generators under conjugation by $g_s$ and $g_s'$ see Appendix \ref{sec:app_SU4}.
The symmetry that governs SSB is the part of ${\cal F}={\cal F}_L\oplus {\cal F}_R$ with $z_G=-I$
and denoting the corresponding eigenvalues of operators by $S_L$ and $S_R$,
we are therefore interested in the ${\cal F}$-eigenvalue $S=S_L\cdot S_R$.
Heavy gauge bosons are represented by the $S$-odd operators $Z,Z^{\pm}$.
The photon $\g$ and the $C=+1$ state ${\overline \g}$, also contained in principle in the spectrum of these operators, if present,
should appear as massless states.
In the table below we summarize the relevant operators of the left boundary, 
their global quantum numbers and the states that they may represent.
\be
\begin{tabular}{|c|c|c|c|c|c|c|c|c|}
\hline 
$O$ & $J$ & $P$ &$C$ & $S_L$ & $S_R$ & $S$ & $CP$ & ${\rm state}$  \\
\hline \hline   
${\rm Re}\, \tr(P_L) $  & 0 & $ + $ & $ + $ & $-$ & $-$ & $ + $& $ + $& $ {\rm Higgs}$         \\ \hline
$\tr(Z_L^+) $  & 1 & $ - $ & $ + $ & $+$ & $-$ & $ - $& $ - $& $ {\overline \g},\, {\rm Heavy\, gauge\, bosons}$         \\ \hline
$\tr(Z_L^-) $  & 1 & $ - $ & $ - $ & $+$ & $-$ & $ - $& $ + $& $ \g,\, {\rm Heavy\, gauge\, bosons}$  \\ \hline
$\tr(Z_L) $  & 1 & $ - $ & $ - $ & $-$ & $+$ & $ - $& $ + $&$ \g,\, {\rm Heavy\, gauge\, bosons}$   \\ \hline
\end{tabular}
\ee
The Lie algebra analysis in Appendix A implies that $H$ breaks spontaneously
to a $U(1)$ via the non-perturbative mechanism. 
Specifically, only one linear combination of generators is invariant under conjugation by both $g_s$ and $g_s'$
so that we expect the total breaking pattern
\be
SU(4)  \,\stackrel{g}{\longrightarrow}\, SU(2)\times SU(2)\times U(1)  \,\stackrel{\rm SSB}{\longrightarrow}\, U(1)\, .
\ee
The natural question is if one can
deduce the existence and perhaps a possible prediction for the value of a Weinberg angle.
In the continuum, since SSB proceeds due to the presence of a local vev,
the Lie algebra contains this information. On the lattice however this is not straightforward.
The reason for the obstruction to connect the Lie algebra
picture with the lattice is partially because on the lattice, by symmetry arguments only, there is no way to tell
how many and which are the physical scalars.
The Polyakov loop operator in its continuum limit gives the sum of all
orbifold-even scalars squared (8 scalars in the $SU(4)$ model). We know that
they can not be all physical since several generators break, however the observable
treats all continuum scalars, physical and non-physical democratically.
This means that quantities like the Weinberg angle have a dynamical origin and can
be determined for example by Monte Carlo simulations.

\section{Non-Perturbative Gauge-Higgs Unification \label{sec:NPGHU}}

The scalar Polyakov Loop in the continuum limit
contains the even under the orbifold projection fields $A_5^{\hat a}$ (for $SU(2)$ these would be $A_5^{1,2}$)
on the boundaries, in the perturbative approach identified with the Higgs field of the four-dimensional effective theory.
When one of these scalars is shifted by $v$, the 1-loop Coleman-Weinberg-Hosotani potential plays the role of
the Higgs potential. Perturbatively this potential does not break any symmetry in the pure gauge theory.
Non-perturbatively we saw on the other hand that the deconfined phase should be Higgs and this has been explicitly
verified by various methods for the $SU(2)$ model. Apparently a mechanism of spontaneous symmetry breaking
is at work, to which perturbation theory (at least at 1-loop) is blind.
For this reason, we call this mechanism "Non-perturbative Gauge-Higgs Unification", NPGHU for short.

In order to see why the perturbative analysis of SSB in the pure gauge case leads to different conclusions,
let us take for concreteness $G=SU(N)$ and try to see if the Higgs mechanism in NPGHU can be interpreted as the shift
\be
A_5 \longrightarrow A_5 + v\label{shift}
\ee
as an attempt to connect to perturbation theory would suggest. Then, since
\be
e^{i (A^A_5T^A+v^AT^A) } = e^{i{A'}^{A}_5T^A}\, e^{i {v'}^A T^A}
\ee
we can introduce such a vev by the shift
\be
{\cal U}(n_5) \longrightarrow {\cal U}(n_5)\, g_v \,.\label{Ushift}
\ee
This shift changes the line $l$ into
\be\label{eq:lv}
l_v = {\cal U}(0)\, g_v\, {\cal U}(1)\, g_v \cdots {\cal U}(N_5-1)\, g_v \,.
\ee
The matrix $g_v$ is a constant $G$-element and it is either in the center of $G$ or not.
Since we are discussing a non-perturbative
mechanism triggered by the breaking of the generalized Weyl group, we would like to see
if the shift of the gauge field by such an element $g_v$ can be interpreted as a stick transformation.
If $g_v$ is in the center, in a perturbative treatment it 
could not trigger SSB, because
$g_v$ commutes in particular with all the algebra generators. 
This is consistently reflected by the fact that $g_v$ cancels from the
Polyakov lines $P_{L(R)}$.
If on the other hand $g_v$ is not in the center of $G$, the vev can not be gauged away from the bulk links
so the shift in \eq{shift} again can not be related to a stick transformation.
To see this we note that under gauge transformation the line $l_v$ in 
\eq{eq:lv} transforms as
\be\label{eq:lvgt}
l_v\,\longrightarrow\,
\Omega(0)\,{\cal U}(0)\,g_v(1)\, {\cal U}(1)\, g_v(2) \cdots 
{\cal U}(N_5-1)\,\Omega(N_5)^\dagger\, g_v \,,
\ee
with $g_v(n_5)=\Omega(n_5)^\dagger\,g_v\,\Omega(n_5)$, $n_5=1,2,\ldots,N_5-1$. 
In order to interpret \eq{eq:lvgt} as a stick symmetry transformation 
$S_R$ in ${\cal F}_R$ with the transformation $l\longrightarrow l\,g_s$ 
(cf. Appendix~\ref{sec:app_calF}), we would need $g_v(n_5)=I$, which implies
$g_v=I$.
Therefore we conclude that a stick symmetry transformation is not equivalent to 
introducing a vev in the scalar Polyakov Loop.
Note that the latter would be the Wilson line breaking mechanism typically employed
in string theory and string inspired models in order to reduce the effective gauge symmetry
in four dimensions: the surviving gauge symmetry is generally determined by the algebra generators of $H$ that
commute with the Wilson line \cite{GSW}. In some cases, in order to declare SSB, the dynamics should harmonize itself with the symmetry argument.
When there are two or more extra dimensions available this means that the tree level potential
of the four-dimensional effective theory should have the proper structure to trigger the expected SSB.
If there is only one extra dimension on the other hand then the scalar potential
vanishes at the classical level but a non-trivial scalar potential for the phase of the Wilson line
may develop at the quantum level. This is now the potential that should trigger the expected SSB, except that
in the absence of fermions it turns out to respect the $H$ symmetry. When fermions are added, SSB can be achieved
and this is the typical context of the Hosotani mechanism in continuum GHU models.
All the above seem to point to NPGHU being a mechanism distinct from other known SSB mechanisms in higher dimensions.
It could be of course that when fermions are introduced for example, the Hosotani mechanism in its lattice version \cite{Cossu} will add to it and one finally
could have a combined mechanism of SSB. Despite the fact that we do not see at the moment if and how fermions will modify
our symmetry-order parameter arguments, a combined SSB picture should not be excluded as a possibility.

Furthermore, in combination with the above discussion, our general analysis suggests that NPGHU is a non-perturbative effect.
The natural question that arises is if it can be advocated as the origin of the Higgs mechanism
in the Standard Model. We leave the possible phenomenological obstructions aside and
discuss only the core of the mechanism. For this, it is sufficient to consider again the $SU(2)$ model for
which we have a sizable amount of information.
In this model we call the massive boundary $U(1)$ gauge boson the $Z$ with mass $m_Z$ and we denote the mass of the Higgs by $m_H$.
We also denote the physical size of the extra dimension by $R$.
The mechanism in this case has been
verified by Monte Carlo and Mean-Field methods, and the latter could shed some more light on its nature.
In \cite{OrbifoldMF1,OrbifoldMF2} we argued that the lattice orbifold is essentially like a relativistic, bosonic superconductor.
This is consistent with the fact that it is a non-perturbative effect. Furthermore, on the anisotropic lattice
there is a regime on the phase diagram where the system reduces dimensionally without the fifth dimension
becoming small. Dimensional reduction occurs instead via the Fu and Nielsen localization mechanism \cite{FuNielsen}.
According to this mechanism, the four-dimensional hyperplanes are weakly coupled while the fifth dimension is strongly coupled (i.e. $\b_4 > \b_5$).
This implies that physics on the orbifold boundaries can be described by a four-dimensional effective action that can be treated perturbatively.
From the superconductor point of view, this would be the Landau functional, i.e. the effective action for
the gauge-scalar system, evaluated on the boundaries.

The precise determination of this effective action is beyond the scope of the present paper, however we can
already extract its general form. Going back to general $SU(N)$, it is expected to be the effective action of the 
order parameter for SSB, say of $Z_k$. It must also be a scalar. 
Then it will have the general form
\bea
{\cal L}_{L} &=& c_1 \sum_k \tr(Z_kZ_k) + c_2 \sum_k \tr(Z_k)\tr(Z_k) 
+ c_3 \sum_{k,l} \tr(Z_kZ_k Z_lZ_l)\nonumber\\
&+& c_4 \sum_k \tr(Z_kZ_k) \sum_l \tr(Z_l)\tr(Z_l) +\cdots
\eea
with the coefficients $c_1, c_2, c_3, c_4,\cdots$ to be determined. 
For concreteness let us consider the vector boson operator defined in \eq{eq:ZL} (dropping the $L$ subscript for clarity)
\be
Z_k = g {\cal T}_k(n) \Phi(n+{\hat k}){\cal T}^\dagger(n) \Phi(n)
\ee
and the expansions in the lattice spacing
\bea
{\cal T}_k &=& e^{aA_k} = 1 + aA_k +\frac{1}{2}a^2A_k^2+\cdots\nonumber\\
\Phi(n+{\hat k}) &=& \Phi(n) + a\, \partial_k \Phi + \frac{1}{2}a^2 \partial_k^2 \Phi+\cdots \,,
\eea
where $A_k^\dagger=-A_k$, $[A_k,g]=0$ and $\{\Ph,g\} = 0$.
Let us define the covariant derivative
\be
{\cal D}_k = \partial_k + 4 A_k
\ee
and the dimensionless\footnote{
A dimensionful Higgs field can be defined from \eq{eq:clPhi}.}
Higgs field $H$ (not to be confused with the boundary gauge symmetry for which we use the same letter)
\be
H = \Phi^2=-\Phi^\dagger\Phi\, .\label{H}
\ee
We obtain
\bea
\tr(Z_k) &=& \frac{1}{2} a\, \tr( g {\cal D}_k H ) + {\rm O}(a^2)
\eea
and
\be
\tr(Z_k Z_k) = \eta\,\tr(H^2) + a\,\eta\, \tr( H \partial_k H ) + {\rm O}(a^2)
\ee
with the sign $\eta$ defined in \eq{eq:etag}.
The effective action now for $Z_k$ then takes the form
\bea\label{eq:Leff}
{\cal L}_{L} &=& 
3\,c_1\, \eta\, \tr(H^2) +
9\, c_3\, \eta^2\,\tr(H^4) \nonumber \\
& + &
a\,\left[ c_1\,\eta \sum_k \tr( H \partial_k H ) +
3\,c_3 \sum_k \tr( H^2 \partial_k H^2 ) \right] + {\rm O}(a^2)
\label{Landau}
\eea
The terms which are O(1) in the lattice spacing in the first line of
\eq{eq:Leff} build up a Higgs potential $V$ for $H$,
to be compared with $V = -\m^2\, \tr(H^2) + \l\,\tr(H^4)$.
It is easy to see that all terms in the potential contain an even number of $g$-insertions
which then annihilate yielding $\pm$ signs.
The coefficients $c_1,c_2,c_3, \cdots$ can be computed numerically by Monte Carlo methods 
or analytically in some approximation scheme, like the mean-field expansion.
We will postpone their computation for the near future.
Notice that for $SU(2k+1)$, ${\bf -1}$ is not a group element so $\eta=1$, while for groups $SU(2k)$, ${\bf -1}$
is always a group element so $\eta=-1$.
Therefore, for $SU(2k)$, the reason for the opposite relative sign in the potential could be that in the quadratic term there are 
two $g$-insertions and in the quartic four $g$-insertions, and that $g^2=\eta=-1$.
Then if $c_1c_3>0$ we have a mexican hat potential.
One observation is that SSB is signaled in the effective action by a vev for the field $H$,
a non-local operator defined by Eqs. (\ref{PL}), (\ref{Phi}) and (\ref{H}).
As such, it can not be represented by the local field $A_5$ taking a vev in an action
with a finite number of terms. 
Another observation is that an effective action of the form \eq{Landau} would have not been possible to obtain
from the effective action of other observables. For example, the plaquette effective action would have not
yielded the potential $V$ because there is no $F_{ij}$ term in 5d, with $i,j$ extra dimensional indices. 
On the other hand, above \eq{Phi} we have stated that a possibility for a scalar operator is essentially $\tr(H)$.
Indeed, its exponential time decay determines the scalar mass spectrum \cite{OrbifoldMC}.
As the ground state in the scalar sector is massive everywhere in the deconfined phase, $H$ has a non-zero expectation value.
A simple calculation now gives $\tr(H)=4\,\tr[(P-P^\dagger)^2]$ which implies that
$P$ can not have the form ${\rm diag}(1,\cdots,1)$, that is, it has non-degenerate eigenvalues. 
Conversely, non-degenerate eigenvalues of $P$ imply a non-zero scalar mass.
These arguments can be actually transferred identically on the fully periodic system
(i.e. without the orbifold boundary conditions), where we know that (in the pure gauge theory) SSB is absent.
All this can be summarized by the statement that it is not clear whether the scalar Polyakov Loop
is the appropriate order parameter for SSB, that role played by the vector Polyakov Loop,
in agreement with our symmetry argument.

In \cite{OrbifoldMF2}, trajectories on the phase diagram along which $m_HR$ and $m_ZR$ 
are constant were constructed for the $SU(2)$ model of Sect. \ref{sec:SU2} .
These Lines of Constant Physics (LCPs) demonstrate the stability of the Higgs mass against quantum fluctuations, at least
in the context of the Mean-Field expansion (work for the Monte Carlo version of these lines is in progress).
A similar question arises in superconductors where one could ask why the effective pole mass of the
Higgs-Anderson field originating from the Cooper pairs is stable under quantum corrections.
Even though in that case there is a natural cut-off scale associated with the size of an atom and one
could argue that even if there is a power dependence of the field's mass on the cut-off, it does not generate a hierarchy problem,
the question in principle remains. One could have a low cut-off and a power law cut-off dependence canceling mechanism at work nevertheless.
We are not aware of such computations regarding superconductors but we know that in the Mean-Field construction
the Higgs mass remains stable across a huge range of the values of the lattice spacing.
Therefore, some kind of cancellation mechanism must be at work.
A possible further hint is the fact that irrespectively of SSB, perturbation theory tells us that $m_HR$ is stable
at one and perhaps even at higher loops. Now given the fact that there is a well defined (if tedious) way to take the perturbative
limit of the Mean-Field expressions for $m_HR$ and the fact that the Mean-Field at each order represents a resummation of an
infinite number of perturbative diagrams, we see two possibilities:
either $m_HR$ remains constant everywhere on the phase diagram, a possibility that can be
dismissed rather easily based on the non-renormalizability of the underlying gauge theory or by simply
looking at Monte Carlo data, or there is a cancelation mechanism from the point of view of the four-dimensional
effective boundary theory.
In real life superconductors the stability of the scalar mass may be simply a direct consequence of the field's fermionic origin but it could also be
that there is something new to be understood there and that this knowledge could be perhaps transferred to our orbifold lattices.

\section{Conclusions}

We argued that spontaneous symmetry breaking in extra dimensional orbifold lattice (pure) gauge theories can be interpreted 
mathematically as the system's spontaneous response to the orbifold projection of becoming sensitive to its generalized Weyl group.
Another, physical point of view sees it as a phenomenon of relativistic, bosonic superconductivity, triggered by the breaking of 
translational invariance in the fifth dimension and the appearance of an effective Higgs field due to the orbifold projections.
It is a non-perturbative mechanism of Gauge-Higgs Unification to which
perturbation theory seems to be blind, called NPGHU in this work.
We have examined mainly models with original $SU(N)$ symmetry.
Realistic model building could involve of course other gauge groups including also product groups.

{\bf Acknowledgments.} We thank H. B. Nielsen and C. Timm for discussions.
FK thanks CERN for hospitality.
This project is funded by the Deutsche Forschungsgemeinschaft (DFG) under
contract KN 947/1-2.
The Monte Carlo simulations were carried out on the
Cheops supercomputer at the RRZK computing centre of the University of Cologne,
which we thank for support.

\begin{appendix}

\section{$SU(4)$ conjugations \label{sec:app_SU4}}

In this Appendix we list the conjugations of the $SU(4)$ generators in the fundamental representation
by $g_s$ and $g_s'$ in \eq{stick4}. The (unnormalized) generators are
\be
H_1 = \begin{pmatrix} 
1 & 0 & 0 & 0 \cr 
0 & -1 & 0 & 0 \cr
0 & 0 & 0 & 0 \cr
0 & 0 & 0 & 0 \cr
\end{pmatrix} \hskip 1cm
H_2 = \begin{pmatrix} 
1 & 0 & 0 & 0 \cr 
0 & 1 & 0 & 0 \cr
0 & 0 & -2 & 0 \cr
0 & 0 & 0 & 0 \cr
\end{pmatrix} 
\hskip 1cm
H_3 = \begin{pmatrix} 
1 & 0 & 0 & 0 \cr 
0 & 1 & 0 & 0 \cr
0 & 0 & 1 & 0 \cr
0 & 0 & 0 & -3 \cr
\end{pmatrix} 
\ee
\be
T^1 = \begin{pmatrix} 
0 & 1 & 0 & 0 \cr 
1 & 0 & 0 & 0 \cr
0 & 0 & 0 & 0 \cr
0 & 0 & 0 & 0 \cr
\end{pmatrix} 
\hskip 1cm
T^2 = \begin{pmatrix} 
0 & -i & 0 & 0 \cr 
i & 0 & 0 & 0 \cr
0 & 0 & 0 & 0 \cr
0 & 0 & 0 & 0 \cr
\end{pmatrix} \hskip 1cm
\ee
\be
T^3 = \begin{pmatrix} 
0 & 0 & 1 & 0 \cr 
0 & 0 & 0 & 0 \cr
1 & 0 & 0 & 0 \cr
0 & 0 & 0 & 0 \cr
\end{pmatrix} 
\hskip 1cm
T^4 = \begin{pmatrix} 
0 & 0 & -i & 0 \cr 
0 & 0 & 0 & 0 \cr
i & 0 & 0 & 0 \cr
0 & 0 & 0 & 0 \cr
\end{pmatrix} \hskip 1cm
\ee
\be
T^5 = \begin{pmatrix} 
0 & 0 & 0 & 0 \cr 
0 & 0 & 1 & 0 \cr
0 & 1 & 0 & 0 \cr
0 & 0 & 0 & 0 \cr
\end{pmatrix} 
\hskip 1cm
T^6 = \begin{pmatrix} 
0 & 0 & 0 & 0 \cr 
0 & 0 & -i & 0 \cr
0 & i & 0 & 0 \cr
0 & 0 & 0 & 0 \cr
\end{pmatrix}
\ee
\be
T^7 = \begin{pmatrix} 
0 & 0 & 0 & 1 \cr 
0 & 0 & 0 & 0 \cr
0 & 0 & 0 & 0 \cr
1 & 0 & 0 & 0 \cr
\end{pmatrix} 
\hskip 1cm
T^8 = \begin{pmatrix} 
0 & 0 & 0 & -i \cr 
0 & 0 & 0 & 0 \cr
0 & 0 & 0 & 0 \cr
i & 0 & 0 & 0 \cr
\end{pmatrix}
\ee
\be
T^9 = \begin{pmatrix} 
0 & 0 & 0 & 0 \cr 
0 & 0 & 0 & 1 \cr
0 & 0 & 0 & 0 \cr
0 & 1 & 0 & 0 \cr
\end{pmatrix} 
\hskip 1cm
T^{10} = \begin{pmatrix} 
0 & 0 & 0 & 0 \cr 
0 & 0 & 0 & -i \cr
0 & 0 & 0 & 0 \cr
0 & i & 0 & 0 \cr
\end{pmatrix}
\ee
\be
T^{11} = \begin{pmatrix} 
0 & 0 & 0 & 0 \cr 
0 & 0 & 0 & 0 \cr
0 & 0 & 0 & 1 \cr
0 & 0 & 1 & 0 \cr
\end{pmatrix} 
\hskip 1cm
T^{12} = \begin{pmatrix} 
0 & 0 & 0 & 0 \cr 
0 & 0 & 0 & 0 \cr
0 & 0 & 0 & -i \cr
0 & 0 & i & 0 \cr
\end{pmatrix}
\ee
and their conjugations
($T^a$ are the $SU(2)\times SU(2)\times U(1)$ generators and $T^{\hat a}$ are the odd under the orbifold generators)
\hskip 2cm
\begin{center}
\begin{tabular}{|c|c|c|}
\hline 
$ O $ & $g_s^\dagger O g_s $ & $g_s^{'-1} O g'_s $ \\
\hline \hline   
$ T^1 + T^{11}$  & $ + $ & $ + $ \\ \hline
$ T^1 - T^{11}$  & $ - $ & $ -$ \\ \hline
$ T^2 + T^{12} $  & $ -$ & $ + $ \\ \hline
$ T^2 - T^{12} $  & $ + $ & $ - $ \\ \hline
$ 1/3(H_3+2H_2) $  & $ - $ & $ - $ \\ \hline
$ H_1+1/3(H_3-H_2) $  & $ -$ & $ + $ \\ \hline
$  H_1-1/3(H_3-H_2) $  & $ + $ & $ - $ \\ \hline
\end{tabular}
\end{center}
The quick rule is that conjugation of a generator by $g_s$
amounts to reflecting the generator with respect to its diagonal,
and then reflecting it once more around its minor diagonal
while conjugation by $g_s'$ simply interchanges the two $SU(2)$ blocks
(this also proves that conjugating an $SU(2)\times SU(2)\times U(1)$
element by $g_s$ or $g_s'$ leaves the element in the group).

\section{Transformations of lattice operators \label{sec:app_transf}}

In this Appendix,
we discuss in detail the transformation properties of the lattice operators
introduced in \sect{sec:orb} under parity $P$, charge conjugation $C$ and
the fixed point symmetry ${\cal F}$.
Finally, their expressions in the classical continuum limit are presented,
which exhibit their spin $J$ quantum number.

In the operators $Z^\pm$ in \eq{eq:Zpm}, the field strength tensor appears.
A symmetric definition of the field strength tensor is given in 
\cite{Luscher:1998pe} using the sum $Q_{MN}(n)$ of four plaquettes in
directions $M$ and $N$ with the same orientation 
(the first link in each plaquette is always pointing towards the point $n$)
\begin{eqnarray}\label{eq:fieldstrength}
Q_{MN}(n) & = & U_M(n)\,U_N(n+\hat{M})\,U_M^\dagger(n+\hat{N})\,U_N^\dagger(n)
\nonumber \\
&& + U_N(n)\,U_M^\dagger(n-\hat{M}+\hat{N})\,U_N^\dagger(n-\hat{M})\,U_M(n-\hat{M})
\nonumber \\
&& + U_M^\dagger(n-\hat{M})\,U_N^\dagger(n-\hat{M}-\hat{N})\,U_M(n-\hat{M}-\hat{N})\,U_N(n-\hat{N}) \nonumber\\
&& + U_N^\dagger(n-\hat{N})\,U_M(n-\hat{N})\,U_N(n+\hat{M}-\hat{N})\,U_M^\dagger(n) \,.
\end{eqnarray}
The anti-Hermitian field strength tensor is given by
\begin{equation}\label{eq:F}
\hat{F}_{MN} = \frac{1}{8a^2}[Q_{MN}(n)-Q_{MN}^\dagger(n)] \,.
\end{equation}
A special case on the orbifold are the expressions for $Q_{k5}$ at $n_5=0$ and
$n_5=N_5$, where only two plaquettes (the ones contained in the fundamental
domain of the orbifold) in \eq{eq:fieldstrength} are used.

\subsection{Parity $P$ \label{sec:app_P}}

The tree-dimensional space reflection or parity $P$ acts on lattice 
coordinates as
\begin{equation}\label{eq:Pn}
P\,n = n_P \,,\quad n_P = (n_0,-\vec{n},n_5)
\end{equation}
and on gauge links as
\begin{equation}\label{eq:PU}
P\,U(n,k) = U^\dagger(n_P-\hat{k},k) \,,\quad
P\,U(n,M) = U(n_P,M) \,,\; (M=0,5) \,.
\end{equation}
It is easy to check that the field strength tensor \eq{eq:F} 
transforms under parity as
$P\,F_{kl}(n) = F_{kl}(n_P)$ and $P\,F_{k5}(n) = -F_{k5}(n_P)$.
The lattice operators introduced in \sect{sec:orb} transform
as\footnote{
We omit the subscripts $L(R)$ when the operators
on the left and right boundaries have the same transformations.}
\begin{eqnarray}
P(n) & \stackrel{P}{\longrightarrow} & P(n_P) \,,\nonumber\\
\Phi(n) & \stackrel{P}{\longrightarrow} & \Phi(n_P) \,,\nonumber\\
\tr[Z_{k}(n)] & \stackrel{P}{\longrightarrow} & -\tr[Z_{k}(n_P-\hat{k})] 
\,,\nonumber\\
Z^\pm_{k}(n) & \stackrel{P}{\longrightarrow} & -Z^\pm_{k}(n_P)
\,.\nonumber
\end{eqnarray}
After the sum over the spatial coordinates $\vec{n}$ is taken
to project to zero spatial momentum $\vec{p}=0$, the operators
$\tr(Z_{k})$ and $\tr(Z_{k}^\pm)$ have parity $P=-1$.

\subsection{Charge conjugation $C$ \label{sec:app_C}}

The charge conjugation $C$ acts on the lattice as complex conjugation
of the gauge links
\begin{equation}\label{eq:C}
C\,U(n,M) = U^*(n,M) \,.
\end{equation}
Under charge conjugation the lattice operators introduced in \sect{sec:orb} 
transform as
\begin{eqnarray}
P(n) & \stackrel{C}{\longrightarrow} & P^*(n) \,,\nonumber\\
\Phi(n) & \stackrel{C}{\longrightarrow} & \eta\Phi^*(n) \,,\nonumber\\
\tr[Z_{k}(n)] & \stackrel{C}{\longrightarrow} & -\tr[Z_{k}(n)] \,,\nonumber\\
\tr[Z^\pm_{k}(n)] & \stackrel{C}{\longrightarrow} & \pm \tr[Z^\pm_{k}(n)]
\,.\nonumber
\end{eqnarray}
Here we use that $g=g^T$ (since $g=\exp(-2\pi i \vec{V}\cdot\vec{H})$ 
\cite{GaugeOrb2} and
the Cartan generators are symmetric $H_i^T = H_i$) and
\begin{equation}\label{eq:etag}
g^* = \eta\,g \,,\quad \eta=\pm1 \,,
\end{equation}
which means $g^2=\eta\,I$.
Note that charge conjugation
is a good quantum number for the orbifold, since if $U = g U g^{-1}$
then using \eq{eq:etag} it follows $U^* = g U^* g^{-1}$, i.e. if $U$ is
projected then also $U^*$ is.

\subsection{Fixed point symmetry \label{sec:app_calF}}

The requirement of a definite transformation under the fixed point symmetry
restricts the matrix $z_G$ in \eq{eq:zG} to be $\pm I$.
Clearly, among simple unitary groups,  the case $-I$ is possible only for $G=SU(2n)$.
The transformations of lattice operators
under the fixed point symmetries ${\cal F}_L$ defined in \eq{eq:calFL}
are
\bea
l   & \longrightarrow & g_F^{-1}\,l \,,\nonumber
\\
P_L & \longrightarrow & z_G\,g_F^{-1}\,P_L\,g_F \,,\nonumber
\\
P_R & \longrightarrow & z_G\,P_R \,,\nonumber
\\
\Phi_L & \longrightarrow & g_F^{-1}\,\Phi_L\,g_F \,,\nonumber
\\
\Phi_R & \longrightarrow & z_G\,\Phi_R \,,\nonumber
\\
Z_{Lk} & \longrightarrow & z_G\,g_F^{-1}\,Z_{Lk}\,g_F \,,\nonumber
\\
Z_{Rk} & \longrightarrow & Z_{Rk} \,,\nonumber
\\
Z^\pm_{Lk} & \longrightarrow & g_F^{-1}\,Z^\pm_{Lk}\,g_F \,,\nonumber
\\
Z^\pm_{Rk} & \longrightarrow & z_G\,Z^\pm_{Rk} \,.\nonumber
\eea
Under under the symmetries ${\cal F}_R$ defined in \eq{eq:calFR}
the transformations are
\bea
l   & \longrightarrow & l\,g_F \,,\nonumber
\\
P_L & \longrightarrow & z_G\,P_L \,,\nonumber 
\\
P_R & \longrightarrow & z_G\,g_F^{-1}\,P_R\,g_F \,,\nonumber
\\
\Phi_L & \longrightarrow & z_G\,\Phi_L \,,\nonumber
\\
\Phi_R & \longrightarrow & g_F^{-1}\,\Phi_R\,g_F \,,\nonumber
\\
Z_{Lk} & \longrightarrow & Z_{Lk} \,,\nonumber
\\
Z_{Rk} & \longrightarrow & z_G\,g_F^{-1}\,Z_{Rk}\,g_F \,,\nonumber
\\
Z^\pm_{Lk} & \longrightarrow & z_G\,Z^\pm_{Lk} \,,\nonumber
\\
Z^\pm_{Rk} & \longrightarrow & g_F^{-1}\,Z^\pm_{Rk}\,g_F \,.\nonumber
\eea

\subsection{Classical continuum limit \label{sec:app_cl}}

In terms of the anti-hermitian linear combination $s$
of the scalars contained in $A_5$, we have the classical continuum limits
\bea
l &=& I+as+\frac{1}{2}a^2 s^2 + {\rm O}(a^3) \nonumber\\
P &=& I\pm a (s-gsg^\dagger) +\frac{1}{2}a^2(gs^2g^\dagger + s^2 -2sgsg^\dagger)+
{\rm O}(a^3) \nonumber\\
\Phi &=& \pm 4a [s,g]+2a^2[gsg^\dagger,s]g+{\rm O}(a^3) \,,\label{eq:clPhi}
\eea
where the upper (lower) sign refers to the operators on the left (right) 
boundary.
By computing the traces one finds for example that
\be
\tr(P) = \tr(I) + 4 \sum_{\hat a} (c_{\hat a}A_5^{\hat a})^2+\cdots
\ee
The continuum limit of the gauge boson operators $Z_{k}$ defined
in \eq{eq:ZL} and \eq{eq:ZR}
is a covariant derivative of the Higgs field \cite{OrbifoldMC}
\bea
\tr[Z_k(x)] &=& a\tr \left[g\,\F(x)\,(\partial_k+2A_k(x))\F(x)\right] + {\rm O}(a^2)\nonumber\\
&=& 32\,a^3\,\eta\,\left\{ 
\tr \left[ (\partial_k s) [g,s] \right] + 
2\,\tr \left[  A_k [s,gs] \right] \right\} + {\rm O}(a^4) \,.
\eea
The continuum limit of the gauge boson operators $Z_{L(R)k}^\pm$ defined
in \eq{eq:Zpm} is
\bea
\tr(Z_k^-) &=& \pm 4\,a^5\,F^a_{12}\,F^{\hat a}_{k5} \tr \left( [s,g] [T^a,T^{\hat a}] \right)+{\rm O}(a^6) \nonumber\\
\tr(Z_k^+) &=& \pm 4\,a^5\,F^a_{12}\,F^{\hat a}_{k5} \tr \left( [s,g] \{T^a,T^{\hat a}\} \right)+{\rm O}(a^6)
\eea
We have used standard notation by which the $G$ Lie algebra index even under
the orbifold projection is $a$ ($g\,T^a\,g^{-1}=T^a$) and the odd is ${\hat a}$ ($g\,T^{\hat a}\,g^{-1}=-T^{\hat a}$).

\end{appendix}



\begin{thebibliography}{99}

\bibitem{GHU}
N.S. Manton, Nucl. Phys. {\bf B158} (1979), 141.
Y. Hosotani, Phys. Lett. {\bf B129} (1983), 193. 

\bibitem{StringOrb}
L.J. Dixon, J.A. Harvey, C. Vafa, E. Witten, Nucl. Phys. {\bf B261} (1985) 678. 
L.J. Dixon, J.A. Harvey, C. Vafa, E. Witten, Nucl. Phys. {\bf B274} (1986) 285.

\bibitem{GaugeOrb1}
E.A. Mirabelli, M. Peskin,  Phys. Rev. {\bf D58} (1998) 065002.
A. Pomarol, M. Quiros,  Phys. Lett. {\bf B438} (1998) 255.
Y. Kawamura, Prog. Theor. Phys. {\bf 103} (2000) 613.
A. Hebecker, J. March-Russell, Nucl. Phys. {\bf B613} (2001) 3. 

\bibitem{GaugeOrb2}
A. Hebecker, J. March-Russell, Nucl. Phys. {\bf B625} (2002) 128. 

\bibitem{TorusMC2}
S. Ejiri, J. Kubo, M. Murata, Phys. Rev. {\bf D62} (2000) 105025.
P. de Forcrand, A. Kurkela, M. Panero, JHEP {\bf 1006} (2010) 050.
K. Farakos, S. Vrentzos, Nucl. Phys. {\bf B862} (2012) 633.
L. Del Debbio, A. Hart, E. Rinaldi, JHEP {\bf 1207} (2012) 178.
F. Knechtli, M. Luz, A. Rago, Nucl. Phys. {\bf B856} (2012) 070.
L. Del Debbio, R.D. Kenway, E. Lambrou, E. Rinaldi, Phys. Lett. {\bf B724} (2013) 133.
L. Del Debbio, R.D. Kenway, E. Lambrou, E. Rinaldi, arXiv:1309.6249 [hep-lat].

\bibitem{Kubo}
M. Kubo, C.S. Lim, H. Yamashita, Mod. Phys. Lett. {\bf A17} (2002) 2249.

\bibitem{Quiros}
G. von Gersdorff, N. Irges, M. Quiros, Nucl. Phys. {\bf B635} (2002) 127.
G. von Gersdorff, N. Irges, M. Quiros, Nucl. Phys. hep-th/0206029.
H.-C. Cheng, K.T. Matchev and M. Schmaltz, Phys. Rev. {\bf D66} (2002) 056006.

\bibitem{'tHooft}
G. 't Hooft, Nucl. Phys. {\bf B79} (1974) 276.

\bibitem{Shigemitsu}
I.-H. Lee, J. Shigemitsu, Nucl. Phys. {\bf B263} (1986) 280.

\bibitem{Hart}
A. Hart, O. Philipsen, J.D. Stack, M. Teper, Phys. Lett. {\bf B396} (1997) 217.

\bibitem{LatOrb1}
N. Irges, F. Knechtli, Nucl. Phys. {\bf B719} (2005) 121.

\bibitem{OrbifoldMC}
N. Irges, F. Knechtli, hep-lat/0604006.
N. Irges, F. Knechtli, Nucl. Phys. {\bf B775} (2007) 283.
N. Irges, F. Knechtli, K. Yoneyama, PoS LATTICE2012 (2012) 056.

\bibitem{Montvay}
I. Montvay, Phys. Lett. {\bf B150} (1985) 441.

\bibitem{Patera}
M. Larouche, F.W.Lemire, J. Patera, J. Phys. A: Math. Theor. {\bf 44} (2011) 415204.

\bibitem{Yaffe}
P.B. Arnold, L.G. Yaffe, Phys. Rev. {\bf D52} (1995) 7208.

\bibitem{Stick}
K. Ishiyama, M. Murata, H. So, K. Takenaga, Prog. Theor. Phys. {\bf 123} (2010) 257.

\bibitem{Shankar}
K. Shankar, Diff. Geom. and Appl. {\bf 14} (2001) 57.

\bibitem{frere}
N. Cosme, J.M. Fr\`{e}re, Phys. Rev. {\bf D69} (2004) 036003.

\bibitem{Elitzur}
S. Elitzur, Phys. Rev. {\bf D12} (1975) 3978.

\bibitem{Knechtli:2014ioa}
F. Knechtli, K. Yoneyama, P. Dziennik and N. Irges,
{\tt arXiv:1402.3491}.

\bibitem{OrbifoldMC2}
P. Dziennik, N. Irges, F. Knechtli, G. Moir, K. Yoneyama, work in progress.

\bibitem{Hasenfratz:2001hp}
A. Hasenfratz and F. Knechtli,
Phys. Rev. {\bf D64} (2001) 034504.

\bibitem{YoneyamaPhD}
K. Yoneyama, 
Ph.D. thesis, \\{\tt http://elpub.bib.uni-wuppertal.de/edocs/dokumente/fbc/physik/diss2014/yoneyama}.

\bibitem{SU3cont}
I. Antoniadis, K. Benakli, Phys. Lett. {\bf B326} (1994) 69.
B. Grzadkowski, J. Wudka, Phys. Rev. Lett. {\bf 97} (2006) 211602.
Y. Adachi, C.S. Lim, N. Maru, Phys. Rev. {\bf D80} (2009) 055025.

\bibitem{GSW}
M.B. Green, J.H. Schwarz, E. Witten, "Superstring Theory Vol 2",
Cambridge Monographs on Mathematical Physics.

\bibitem{Cossu}
G. Cossu, H. Hatanaka, Y. Hosotani and J.-I. Noaki,
arXiv:1309.4198 [hep-lat].
G. Cossu, E. Itou, H. Hatanaka, Y. Hosotani, J-I. Noaki, arXiv:1311.0079 [hep-lat].
K. Kashiva, T. Misumi, JHEP {\bf 1305} (2013) 042.

\bibitem{OrbifoldMF1}
N. Irges, F. Knechtli, K. Yoneyama, Nucl. Phys. {\bf B865} (2012) 541.

\bibitem{OrbifoldMF2}
N. Irges, F. Knechtli, K. Yoneyama, Phys. Lett. {\bf B722} (2013) 378.

\bibitem{FuNielsen}
Y. K. Fu and H. B. Nielsen, Nucl. Phys. {\bf B236} (1984) 167.

\bibitem{Luscher:1998pe}
M. L\"uscher,
{\tt hep-lat/9802029}.


\end{thebibliography}
\end{document}